\documentclass[a4paper, 12pt]{article}
\usepackage[T1]{fontenc}
\usepackage{jheppub,graphicx,epsf,amssymb,amsbsy,amsfonts,amssymb,amsmath, empheq,mathrsfs,appendix,soul}
\usepackage{hyperref}
\def\be{\begin{equation}}
\def\ee{\end{equation}}
\def\bea{\begin{eqnarray}}
\def\eea{\end{eqnarray}}

\def\bg{\bar{g}}
\def\beq{\begin{eqnarray}}\def\eeq{\end{eqnarray}}
\def\ba#1\ea{\begin{align}#1\end{align}}
\def\bg#1\eg{\begin{gather}#1\end{gather}}
\def\bm#1\em{\begin{multline}#1\end{multline}}
\def\bmd#1\emd{\begin{multlined}#1\end{multlined}}

\def\D{\Delta}

\def\({\left(}
\def\){\right)}
\def\[{\left[}
\def\]{\right]}

\def\D{\Delta}

\title{Boundary operators in asymptotically flat space-time}
\author{Shamik Banerjee}
\affiliation{National Institute of Science Education and Research (NISER), Bhubaneswar 752050, Odisha, India \\ Homi Bhabha National Institute, Anushakti Nagar, Mumbai, India-400085}

\emailAdd{ banerjeeshamik.phy@gmail.com}
\abstract{In \cite{Jain:2023fxc} the authors have proposed an interesting framework for studying holography in flat space-time. In this note we explore the relationship between their proposal and the Celestial Holography. In particular, we find that in both the massive and in the massless cases the asymptotic boundary limit of the bulk time-ordered Green's function $G$ is related to the Celestial amplitudes by an integral transformation. In the massless case the integral transformation reduces to the well known \textit{shadow transformation} of the celestial amplitude. Now the relation between the asymptotic limit of $G$ and the celestial amplitudes suggests that in asymptotically flat space-time if the scattering states are described by the conformal primary basis then the boundary operators defined by the extrapolate dictionary of \cite{Jain:2023fxc} are given by the \underline{shadow transformation} of the conformal primary operators living on the celestial sphere. This result refers to the non-contact part of the extrapolated Green's function. There are important contact term contributions which we also discuss in the paper. }
\date{}

\begin{document}
\maketitle
\flushbottom





\section{Introduction}
In a recent paper \cite{Jain:2023fxc} the authors have proposed an interesting framework for studying holography in asymptotically flat space-time. The central object in this proposal is the Boundary Correlation Function $G_{\text{bnd}}$ which is calculated, in analogy with the AdS-CFT correspondence, from an Euclidean path integral with Dirichlet boundary condition. After analytic continuation to Lorentzian signature one gets a relation between $G_{\text{bnd}}$ and the $S$-matrix elements. This relation is non-local and the $S$-matrix element is expressed as an integral over the $G_{\text{bnd}}$ weighed by free particle wave functions. Therefore $G_{\text{bnd}}$ is a smeared form of the $S$-matrix. 

Another interesting result of \cite{Jain:2023fxc} is the "Inverse LSZ Formula" which expresses the asymptotic boundary limit of the bulk time-ordered Green's function $G$ in terms of the $S$-matrix element. It turns out that when all the particles are the massive $G_{\text{bnd}}$ can be obtained from $G$ by differentiation which we describe in more detail later. Therefore $G$ and $G_{\text{bnd}}$ are related locally. In the case of massless particles the relation between $G$ and $G_{bnd}$ is not straightforward however, for special configuration of boundary points the relation between $G_{\text{bnd}}$ and $G$ is again local. 

Let us now briefly discuss another approach to the Flat Space Holography known as the Celestial Holography \cite{Strominger:2017zoo,Pasterski:2021rjz,Donnay:2023mrd,Pasterski:2016qvg,Pasterski:2017kqt,Banerjee:2018gce}. In celestial holography the central objects are the celestial correlation functions or Celestial Amplitudes $\mathcal A$ which are the $S$-matrix elements written in the "Conformal Primary Basis". The celestial correlation functions are supported on the celestial sphere on which the Lorentz group acts as the group of conformal transformations. Under (Lorentz) conformal transformations celestial correlation functions transform as correlation functions of a conformal field theory. It is conjectured that in $D$ dimensional asymptotically flat space-time the celestial correlation functions are computed by a dual $\text{CFT}_{D-2}$ living on the celestial sphere $\mathcal CS^{D-2}$. The dual CFT is very symmetric, especially in $D=4$, and the Ward identities of the symmetries are the same as the soft factorization theorems. 

In this paper we find out the relation between the celestial amplitude $\mathcal A$ and ($G_{\text{bnd}}$) $G$.  \\

\subsection{Notation and Convention}
In this paper we work with the space-time signature $\(+ - - \cdots\)$. We denote the space-time dimension by $D$ and $d=D-2$ is the dimension of the celestial sphere $\mathcal CS^d$. Our convention for scattering amplitude is that all particles are outgoing.

\section{results}
The results can be summarized as follows: 

1) For $n$ massive particles ($G_\text{bnd}$) $G$ can be expressed as a function of $n$ unit time-like vectors $\hat x_i$\footnote{$\hat x_i = \frac{x_i}{\sqrt{x_i^2}}$ as $x_i^2\rightarrow\infty$.} and the celestial sphere $\mathcal CS^{d}$ can be thought of as the boundary of the hyperboloid $\hat x^2 = 1$ with (stereographic) coordinates $\vec w \in R^{d}$. The transformation formula is given by \eqref{main1}
\be
\begin{gathered}
G_{\text{bnd}}(\{\hat x_i,\delta_i\}) \\ 
= \mathcal N \(2 \prod_i \int_{0}^{\infty} d\lambda_i \mu(\lambda_i) \int d^{d}\vec w_i  G_{\D_i}(\hat x_i, \vec w_i)\mathcal{A}(\{\vec w_i, d-\D_i, \delta_i\})\), \  \D = \frac{d}{2} + i \lambda
\end{gathered}
\ee
where $G_{\D}\(\hat x, \vec w\)$ is the scalar Euclidean $AdS_{D-1}$ bulk-boundary propagator, $N$ is a normalization factor and $\delta_i=\pm 1$ depending on whether the $i$-th particle is outgoing or incoming. The measure factor $\mu(\lambda)$ is defined in \eqref{measure}. Note that the prefactor $\mathcal N$ can be written as a product of factors each of which depends on the mass and coordinates of a \textit{single} particle. 

2) For massless particles the transformation law is given by \eqref{massless}
\be\label{massless1}
\begin{gathered}
G(\{x_i\rightarrow\infty,\delta_i\}) = \prod_{i} \frac{1}{(2\pi)^{d}} \int_{-\infty}^{\infty} \frac{d\lambda_i}{2\pi} \Gamma(\D_i) \frac{(-i\delta_i)^{\D_i}}{R_i^{\D_i}} \int d^d \vec w_i \frac{1}{|\vec z_i - \vec w_i|^{2\D_i}} \mathcal{A}\(\{\vec w_i, d- \D_i, \delta_i\}\) \\
= \prod_{i} \frac{1}{2^d\pi^{\frac{d}{2}}} \int_{-\infty}^{\infty} \frac{d\lambda_i}{2\pi} \Gamma\(\frac{d}{2} - \D_i\) \frac{(-i\delta_i)^{\D_i}}{R_i^{\D_i}} \tilde{\mathcal{A}}\(\{\vec z_i, \D_i, \delta_i\}\)
\end{gathered}
\ee
where $\tilde{\mathcal A}$ is the \textit{shadow transformation} of the celestial amplitude $\mathcal A$. In \eqref{massless1} we have parametrized the $x_i$ in terms of the retarded coordinates $(R, u, \vec z )$ and have taken $R\rightarrow\infty$ at fixed $(u, \vec z)$. So $\vec z$ are the (stereographic) coordinates on the celestial sphere $\mathcal CS^{d}$. $G_{\text{bnd}}$ can be computed from $G$ for special choices of configuration of the points $x_i\rightarrow\infty$. 

There are important contact term contribution to the right hand side of \eqref{massless1} which we discuss in detail in section-5. The paper \cite{Jain:2023fxc} deals only with non-contact terms in the boundary correlation functions.

\subsection{Boundary operators and observables in asymptotically flat space-time}
In \cite{Jain:2023fxc} the authors have defined boundary operators $\{O_{I}\}$ in general space-time. The definition is such that in asymptotically $AdS$ space-time $O_I$ reduces to the standard boundary CFT operator. In asymptotically flat space-time $O_{I}$ reduces to the (creation) annihilation operator for scattering states and correlation functions of $O_I$'s give the $S$-matrix elements. If we write $O_{I}$ in the conformal primary basis \cite{Pasterski:2016qvg,Pasterski:2017kqt,Banerjee:2018gce} then we get the celestial correlation functions. In other words, the conformal primary operator $\phi_{\D,\sigma}^{\delta}(\vec z)$ \footnote{Conformal primary operator in the massless case is defined as
\be\nonumber
\phi_{\D,\sigma}^{\delta} = \int_{0}^{\infty} d\omega \omega^{\D-1} a(\delta p, \sigma)
\ee
where $a(\delta p,\sigma)$ is the is the (Fock) momentum space annihilation or creation operator depending on whether $\delta=1$ or $\delta=-1$, $\sigma$ is the spin and $p$ is the null momentum parametrized as, $p= \omega (1+ \vec z^2, 2\vec z, 1-\vec z^2), \  \omega>0$. Under Lorentz transformation which acts on $\mathcal CS^{D-2}$ as conformal transformation, $\phi_{\D,\sigma}^{\delta}(\vec z)$ transforms as a conformal primary operator of weight $\D$ and spin $\sigma$. $\tilde\phi$ is the shadow transformation of this conformal primary operator. Note that in \eqref{shadow} we have scalar particles with $\sigma=0$ but, it has obvious generalization to spinning particles. } is simply $O_I$ written in the conformal primary basis.

However, equation \eqref{massless1} suggests another definition of boundary operators in asymptotically flat space-time. Equation \eqref{massless1} can be written schematically as 
\be\label{shadow}
G(\{x_i\rightarrow\infty,\delta_i\}) \sim \sum_{\{\D_i\}} R^{-\sum_i \D_i} < \tilde\phi_{\D_1}^{\delta_1}(\vec z_1)\cdots  \tilde\phi_{\D_n}^{\delta_n}(\vec z_n)>
\ee
where $\tilde\phi_{\D}^{\delta}$ is the shadow transform of the conformal primary operator $\phi_{d-\D}^{\delta}$. Equation \eqref{shadow} describes the \textit{asymptotic boundary limit} of the \textit{bulk} (time-ordered) Green's function $G$. This suggests that: \\

In asymptotically flat space-time if the scattering states are described by the conformal primary basis then \textit{ the boundary operators defined by the \underline{extrapolate dictionary} of \cite{Jain:2023fxc} are given by the \underline{shadow transformation} of the conformal primary operators living on the celestial sphere.}\\

The above statement is not without caveat. As we discuss in section-5 the extrapolated Green's function has contact term contributions which are not given by the Celestial shadow amplitude. So many things need to be done before we can consider the question of boundary operators to be settled. But this seems to be on the right track and is consistent with the fact that in celestial holography the \textit{boundary stress tensor is the shadow of the subleading soft graviton}\cite{Kapec:2016jld,Kapec:2014opa,He:2017fsb,Banerjee:2022wht,Kapec:2017gsg}. 


Shadow and light transformation have played important roles in celestial holography \cite{Strominger:2021mtt,Adamo:2021lrv,Sharma:2021gcz,Jorstad:2023ajr}. One of the places where it is most useful is in the study of soft symmetries \cite{Strominger:2021mtt}. Typically soft operators have negative integer conformal dimensions but, after shadow transformation, the dimensions become positive. Therefore for many reasons appearance of shadow operators is satisfying. 



\subsection{Comments on the $4$-point function}
In $D=4$ Celestial $4$-point function is a distribution supported on the configuration $z=\bar z$ where $z$ is the cross-ratio of four points\footnote{Here $z_i\in \mathbb C$ is the sterepgraphic coordinate on the celestial sphere $\mathcal CS^2$ and bar denotes complex conjugation.} on $\mathcal CS^2$. This distributional nature is a consequence of the global space-time translation Ward-identity. However, it is clear that once we do the shadow transformation \eqref{massless1}, the resulting amplitude is generically non-zero away from the $z=\bar z$ submanifold. Therefore the shadow $4$-point amplitude is not a distribution. 

This is consistent with the finding in \cite{Jain:2023fxc} that $G$ or $G_{\text{bnd}}$ is smooth away from the singular submanifold $z=\bar z$. They also found that the singularity is of a simple pole type. We leave the analysis of the singularity structure of the shadow amplitude to future work. 

We would like to point out that there are variants \cite{Melton:2023bjw,Melton:2024jyq,Sleight:2023ojm,Iacobacci:2024nhw} of the conventional celestial amplitudes which do not show distributional nature and have the structure of conventional CFT correlation functions. It will be interesting to know the relation of these amplitudes to the boundary correlation functions as defined in \cite{Jain:2023fxc}.

\section{Derivation}

\subsection{Massive Scalar particles}

For massive particles S-matrix elements are (distributions) functions defined on the mass hyperboloid $p_i^2 = m_i^2$. Given this we can define a function of the space time coordinates by
\be\label{ft}
G(\{x_i,\delta_i\}) = \prod_i \int d\mu_{p_i} e^{-i\delta_i p_i\cdot x_i} S(\{\delta_i p_i\}), \ \delta_i = \pm1 
\ee
where $d\mu_{p_i} = \frac{d^{D-1}p_i}{(2\pi)^{D-2} 2\omega_{p_i}}$ is the Lorentz invariant integration measure on the mass hyperboloid, $p^0=\omega_p = \sqrt{\vec p^2 + m^2}$ and $\delta_i = \pm 1$ depending on whether the $i$-th particle is outgoing or incoming. If we take $x_i\rightarrow \infty$ then \eqref{ft} becomes the "inverse LSZ formula" of \cite{Jain:2023fxc} with $G$ now interpreted as the time-ordered (bulk) Green's function. 

Now we write the equation \eqref{ft} in terms of the conformal primary basis. The change of basis is given by \cite{Pasterski:2016qvg,Pasterski:2017kqt}
\be
e^{-i\delta p\cdot x} = 2\int_{0}^{\infty} d\lambda \mu(\lambda) \int d^{d}\vec w G_{d-\D} (\hat p ; \vec w) \phi^{\delta}_{\D}(x; \vec w), \  \D = \frac{d}{2} + i\lambda, \ p = m \hat p
\ee
where 
\be\label{measure}
\mu(\lambda) = \frac{\Gamma(\D)\Gamma(d-\D)}{4\pi^{d+1}\Gamma(\D-\frac{d}{2})\Gamma(\frac{d}{2}-\D)}
\ee
and 
\be
\phi^{\delta}_{\D}(x,\vec w) = \frac{2^{\frac{d}{2}+1} \pi^{\frac{d}{2}}}{(i m)^{\frac{d}{2}}} \frac{(\sqrt{x^2})^{\D-\frac{d}{2}}}{(q(\vec w)\cdot x -i\delta\epsilon)^{\D}} K_{\D - \frac{d}{2}}(m \sqrt{-x^2}), \  \epsilon \rightarrow 0+
\ee
is the conformal primary wave function. We have also introduced a uni null vector 
\be
q(\vec w) = (1+\vec w^2, 2 \vec w, 1- \vec w^2), \  \vec w \in R^{D-2}
\ee
and $G_{\D}(\hat p,\vec w)$ is the scalar Euclidean $\text{AdS}_{D-1}$ bulk-boundary propagator given by\footnote{This is the familiar embedding space form of the scalar AdS bulk-boundary propagator. Note that $\hat p^2 = 1$ and so the tip of the vector $\hat p$ lies on a Euclidean $AdS_{D-1}$. The vector $\vec w \in R^{D-2}$ gives a point on the boundary of the (Poincare patch of) $AdS_{D-1}$.}
\be
 G_{\D} (\hat p, \vec w) = \frac{1}{\(\hat p\cdot q(\vec w)\)^{\D}}
\ee

After substituting everything in \eqref{ft} we get
\be\label{}
\begin{gathered}
G(\{x_i,\delta_i\}) = 2 \prod_i \int d\mu_{p_i} \int_{0}^{\infty} d\lambda_i \mu(\lambda_i) \int d^{d}\vec w_i G_{d-\D_i} (\hat p_i ; \vec w_i) \phi^{\delta_i}_{\D_i}(x_i; \vec w_i) S(\{\delta_i p_i\}) \\ 
= 2 \prod_i m_i^{D-2} \int d\mu_{\hat p_i} \int_{0}^{\infty} d\lambda_i \mu(\lambda_i) \int d^{d}\vec w_i G_{d-\D_i} (\hat p_i ; \vec w_i) \phi^{\delta_i}_{\D_i}(x_i; \vec w_i) S(\{\delta_i p_i\}) \\
= 2 \prod_i m_i^{D-2} \int_{0}^{\infty} d\lambda_i \mu(\lambda_i) \int d^{d}\vec w_i \phi^{\delta_i}_{\D_i}(x_i; \vec w_i) \mathcal{A}(\{\vec w_i,d- \D_i, \delta_i\})
\end{gathered}
\ee
where $\mathcal A(\{\vec w_i, \D_i, \delta_i\})$ is the celestial amplitude defined by \cite{Pasterski:2016qvg,Pasterski:2017kqt}
\be
\mathcal A(\{\vec w_i, \D_i, \delta_i\}) = \int d\mu_{\hat p_i} G_{\D_i}(\hat p_i, \vec w_i) S(\{\delta_i m_i \hat p_i\})
\ee
Note that $\vec w_i\in R^d$ should be interpreted as the stereographic coordinates of the celestial sphere $\mathcal CS^d$.

Let us now take the asymptotic boundary limit in which all the $x_i$'s are large and time-like\footnote{We can also take some of them to be space-like or null with obvious changes in the final formulas.}. In this limit the conformal primary wave function simplifies to
\be
\begin{gathered}
\phi^{\delta}_{\D}(x,\vec w) = \frac{2^{\frac{d}{2}+1} \pi^{\frac{d}{2}}}{(i m)^{\frac{d}{2}}} \frac{(\sqrt{x^2})^{\D-\frac{d}{2}}}{(q(\vec w)\cdot x -i\delta\epsilon)^{\D}} K_{\D - \frac{d}{2}}(m \sqrt{-x^2}) \\
\sim \frac{2^{\frac{d}{2}+1} \pi^{\frac{d}{2}}}{(i m)^{\frac{d}{2}}} \frac{(\sqrt{x^2})^{-\frac{d}{2}}}{(q(\vec w).\hat x -i\delta\epsilon)^{\D}} 
\sqrt{\frac{\pi}{2m\sqrt{-x^2}}} e^{-m\sqrt{-x^2}}, \  \hat x^{\mu} = \frac{x^{\mu}}{\sqrt{x^2}} \\
= \frac{2^{\frac{d}{2}+1} \pi^{\frac{d}{2}}}{(i m)^{\frac{d}{2}}} \frac{(\sqrt{x^2})^{-\frac{d}{2}}}{(q(\vec w)\cdot \hat x -i\delta\epsilon)^{\D}} 
\sqrt{\frac{\pi}{2im\sqrt{x^2}}} e^{-im\sqrt{x^2}}
\end{gathered}
\ee

With this the formula for $G$ becomes,
\be\label{G}
G(\{x_i, \delta_i\}) = F\(\{m_i, \sqrt{x_i^2}\}\) \(2 \prod_i \int_{0}^{\infty} d\lambda_i \mu(\lambda_i) \int d^{d}\vec w_i  \frac{1}{(q(\vec w_i)\cdot\hat x_i -i\delta_i\epsilon)^{\D_i}} \mathcal{A}(\{\vec w_i, d-\D_i, \delta_i\})\) 
\ee

\subsubsection{$G_{\text{bnd}}$}
In the case of massive particles, $G_{\text{bnd}}$ can be obtained from $G$ by differentiating each external leg in the "normal" direction
\be
G_{\text{bnd}} = \prod_i 2i\hat n.\nabla_i G
\ee
Let us now define the normal direction $\hat n$.

Since $x_i$'s are all time-like and (infinitely) large we can introduce the Milne coordinates which cover the future time-like region of the light-cone emanating from the centre of the Minkowski space. This is also known as the hyperbolic foliation of the Minkowski space. 

In these coordinates space-like surfaces are given by the family of hyperboloids
\be
x^2 = \tau^2, \  0< \tau < \infty
\ee

where $\tau$ is the Milne time. In these coordinates Minkowski metric is given by 
\be
ds^2 = d\tau^2 - \tau^2 ds_{H_{D-1}}^2
\ee
where $H_{D}$ is the $D$ dimensional hyperboloid (or Euclidean $AdS_{D}$). 

\cite{Jain:2023fxc} introduced the slab space-time. We take the future space-like boundary of the slab space-time to be the constant $\tau$ surface
\be
\tau = \tau_0 \rightarrow \infty
\ee
and so the $\tau$ direction is the normal direction and one can easily check that
\be
2i\hat n.\nabla = 2i\partial_{\tau}
\ee

Now in these coordinates the unit time-like vector $\hat x$ is given by \cite{Pasterski:2016qvg,Pasterski:2017kqt}
\be
\hat x = \frac{x}{\tau} = \( \frac{1+y^2 + \vec z^2}{2y}, \frac{\vec z}{y}, \frac{1-y^2 - \vec z^2}{2y}\)
\ee
where $(y,\vec z)$ are the Poincare coordinates on the hyperboloid $H_{D-1}$ (or Euclidean $AdS_{D-1}$). 

Substituting everything in \eqref{G} we get 
\be
\begin{gathered}
G(\{x_i,\delta_i\}) \\ = F\(\{m_i, \tau_i \}\) \(2 \prod_i \int_{0}^{\infty} d\lambda_i \mu(\lambda_i) \int d^{d}\vec w_i  \(\frac{y_i}{y_i^2 + \(\vec w_i - \vec z_i\)^2 -i \delta_i\epsilon}\)^{\D_i}\mathcal{A}(\{\vec w_i, d-\D_i, \delta_i\})\) 
\end{gathered}
\ee

We can see that only the prefactor $F$ has $\tau$ dependence and so the $G_{\text{bnd}}$ is given by\footnote{Here we have omitted the $i\epsilon\delta_i$ prescription because is no singularity in the domain of the integration. However, if we want to analytically continue to the region where some of the $x_i$'s are space-like then the $i\epsilon$ prescription becomes important.} 
\be\label{main1}
\boxed{
\begin{gathered}
G_{\text{bnd}}(\{\hat x_i,\delta_i\}) \\ = \(\lim_{\tau_0\rightarrow\infty} F'\(\{m_i, \tau_0 \}\)\) \(2 \prod_i \int_{0}^{\infty} d\lambda_i \mu(\lambda_i) \int d^{d}\vec w_i  \(\frac{y_i}{y_i^2 + \(\vec w_i - \vec z_i\)^2}\)^{\D_i}\mathcal{A}(\{\vec w_i, d-\D_i, \delta_i\})\) \\
= \mathcal N \(2 \prod_i \int_{0}^{\infty} d\lambda_i \mu(\lambda_i) \int d^{d}\vec w_i  G_{\D_i}(\hat x_i, \vec w_i)\mathcal{A}(\{\vec w_i, d-\D_i, \delta_i\})\)
\end{gathered}}
\ee
where prime on $F$ denotes derivative with respect to $\tau$.

Equation \eqref{main1} is the main result in the massive case which relates the boundary correlation function $G_{\text{bnd}}(\{\hat x_i,\delta_i\})$ to the celestial amplitude $\mathcal A(\{\vec w_i, \D_i, \delta_i)$. It shows that the $G_{\text{bnd}}$ is the pull-back of the celestial amplitude to the hyperboloid using the Euclidean $AdS$ bulk-boundary propagator. 

We now show that the relation \eqref{main1} is invertible, i.e, the celestial amplitude can also be written as an integral over the boundary correlation function. In order to do this we use the following identity which holds when the dimension $\D$ takes values in the unitary principal series representation of the Lorentz group $SO(D-1,1)$. This means, $\D = \frac{d}{2} + i\lambda, \lambda\in R$. In this case  
\be\label{inv}
\begin{gathered}
\int_{H_{d+1}} d\mu_{\hat x} G_{\D}(\hat x, \vec w_1) G_{\D'}(\hat x, \vec w') \\
= \frac{1}{2\mu(\lambda)} \delta^d(\vec w - \vec w') \delta(\lambda + \lambda') + 2\pi^{\frac{d}{2}+1} \frac{\Gamma\(\D - \frac{d}{2}\)}{\Gamma(\D)} \frac{1}{|\vec w - \vec w'|^{2\D}} \delta(\lambda - \lambda')
\end{gathered}
\ee
where $\mu(\lambda)$ has been defined in \eqref{measure}. Using \eqref{inv} one can easily check that the inverse of \eqref{main1} is\footnote{Note that the invertibility emerges only in the large $x_i$ limit.}
\be
\boxed{
\mathcal{A}\(\{\vec w_i, \D_i, \delta_i\}\) = {\mathcal N}^{-1} \prod_{i} \int_{H_{d+1}} d\mu_{\hat x_i}G_{\D_i}(\hat x_i, \vec w_i) G_{\text{bnd}}(\{\hat x_i,\delta_i\})}
\ee

\section{Massless scalar particles}

In the massless case we can again start with the basic equation \eqref{ft} with obvious modificatios
\be\label{ft2}
G(\{x_i,\delta_i\}) = \prod_i \int d\mu_{p_i} e^{-i\delta_i p_i\cdot x_i} S(\{\delta_i p_i\}), \ \delta_i = \pm1
\ee
where $d\mu_{p_i} = \frac{d^{D-1}p_i}{(2\pi)^{D-2} 2\omega_{p_i}}$ is the Lorentz invariant integration measure on the cone and $p^0=\omega_p = |\vec p|$.

For massless particles it is useful to parametrise the null momentum as
\be
p^{\mu} = \omega q^{\mu}(\vec w) = \omega (1+\vec w^2, 2 \vec w, 1- \vec w^2), \  \vec w \in R^{D-2}, \ \omega \ge 0
\ee
In this parametrization the integration measure $d\mu_{p}$ on the cone becomes 
\be
d\mu_p = \frac{1}{(2\pi)^{D-2}}2^{D-3} \omega^{D-3} d\omega d^{D-2}\vec w = \frac{1}{2\pi^{d}}\omega^{d-1} d\omega d^{d}\vec w
\ee
We now transform to the conformal primary basis using the relation \cite{Pasterski:2017kqt}
\be
e^{-i\delta \omega q\cdot x} = \int_{-\infty}^{\infty} \frac{d\lambda}{2\pi} \omega^{-\D} \frac{{(-i\delta})^{\D} \Gamma(\D)}{(q(\vec w)\cdot x - i\delta\epsilon)^{\D}}, \  \D = \frac{d}{2} + i\lambda
\ee
Substituting this in \eqref{ft2} we get 
\be\label{massless}
\begin{gathered}
G(\{x_i,\delta_i\}) = 
\prod_{i} \frac{1}{2\pi^{d}} \int_{0}^{\infty} \omega_i^{d-1} d\omega_i \int d^{d}\vec w_i \int_{-\infty}^{\infty} \frac{d\lambda_i}{2\pi} \omega_i^{-\D_i} \frac{{(-i\delta_i})^{\D_i} \Gamma(\D_i)}{(q(\vec w_i)\cdot x_i - i\delta_i\epsilon)^{\D_i}} S(\{\delta_i p_i\}) \\
= \prod_{i} \frac{1}{2\pi^{d}} \int_{-\infty}^{\infty} \frac{d\lambda_i}{2\pi} (-i\delta_i)^{\D_i} \Gamma(\D_i) \int d^d \vec w_i \frac{1}{(q(\vec w_i)\cdot x_i - i\delta_i\epsilon)^{\D_i}} \mathcal{A}\(\{\vec w_i, d- \D_i, \delta_i\}\)
\end{gathered}
\ee
where $\mathcal{A}$ is the massless celestial amplitude defined as \cite{Pasterski:2017kqt}
\be
\mathcal{A}\(\{\vec w_i, \D_i, \delta_i\}\) = \prod_i \int_{0}^{\infty} d\omega_i \omega_i^{\D_i-1} S(\{\delta_i p_i\})
\ee

Let us now extract the asymptotic limit of \eqref{massless} when all the $x_i$'s are taken to the future null infinity. To study this limit it is useful to transfer to the retarded coordinates 
\be
u = t - r , \  -\infty < u < \infty
\ee
In this coordinate system we can write any $x$ as 
\be
x^{\mu} = u (1,\vec 0) + r (1, \vec n), \  \vec n \cdot \vec n = 1,  \  \vec n \in R^{D-1}
\ee
One can further simplify life by writing the null vector $r(1,\vec n)$ as 
\be
r(1,\vec n) = R q^{\mu} (\vec z), \  r = R(1+ {\vec z}^2),  \  \vec z \in R^{D-2}= R^{d}
\ee
for some $\vec z$. In terms of these objects the scalar product $q(\vec w)\cdot x$ becomes
\be
q(\vec w)\cdot x = u (q^0, \vec 0) + R q(\vec w)\cdot q(\vec z) = u q^0 + 2R | \vec w - \vec z|^2
\ee
The null infinity can be approached by letting $R\rightarrow\infty$ at fixed $u$. In this limit we can neglect the term $uq^0$ if we are interested only in the leading order in $R^{-1}$ expansion. So at leading order we get\footnote{Here we have neglected the $i\epsilon\delta_i$ prescription for the sake of clarity.}\footnote{We have defined the shadow transformation following \cite{Dolan:2011dv} such that the square of the shadow operation is an identity transformation.}
\be\label{massless}
\boxed{
\begin{gathered}
G(\{x_i\rightarrow\infty,\delta_i\}) = \prod_{i} \frac{1}{2\pi^{d}} \int_{-\infty}^{\infty} \frac{d\lambda_i}{2\pi} \Gamma(\D_i) \frac{(-i\delta_i)^{\D_i}}{R_i^{\D_i}} \int d^d \vec w_i \frac{1}{|\vec z_i - \vec w_i|^{2\D_i}} \mathcal{A}\(\{\vec w_i, d- \D_i, \delta_i\}\) \\
= \prod_{i} \frac{1}{2\pi^{\frac{d}{2}}} \int_{-\infty}^{\infty} \frac{d\lambda_i}{2\pi} \Gamma\(\frac{d}{2} - \D_i\) \frac{(-i\delta_i)^{\D_i}}{R_i^{\D_i}} \tilde{\mathcal{A}}\(\{\vec z_i, \D_i, \delta_i\}\)
\end{gathered}}
\ee
where $\tilde{\mathcal A}$ is the \textit{shadow transformation} of the celestial amplitude $\mathcal A$. This is our main result. Note that the approximation which leads to \eqref{massless} is valid only when the shadow integral is convergent which we assume to be the case.\footnote{Typically the shadow integral converges when the scaling dimensions lie in a certain domain. Outside the domain the integral can be defined by analytic continuation. The same is true for Celestial amplitudes. Although to start with it is defined only on the unitary principal series but the final answer can be analytically continued to a much larger domain. This analytic continuation plays a very important role in Celestial Holography.}

\section{Contact term contributions}\label{conatct1}
In our derivation of \eqref{massless} we have neglected the contact term contributions\footnote{I thank the referee for pointing this out to me.}. The expression
\be
u q^0 + 2R | \vec w - \vec z|^2 \sim 2R | \vec w - \vec z|^2, \  R\rightarrow\infty
\ee
only when $|\vec z - \vec w| \ne 0$. But, since $\vec w$ is being integrated over we have to take into account contact term contributions. Contact term contributions can be isolated using the following identity \cite{Pasterski:2017kqt, Donnay:2022sdg}
\be
\(\frac{y}{y^2 + |\vec z - \vec w|^2}\)^{\D} \sim \pi^{\frac{d}{2}}\frac{\Gamma\(\D - \frac{d}{2}\)}{\Gamma({\D})} y^{d-\D} \delta^d(\vec z -\vec w) + \frac{y^{\D}}{|\vec z - \vec w|^{2\D}}, \  y\rightarrow 0
\ee
Cancelling $y^{\D}$ from both sides we get
\be
\(\frac{1}{y^2 + |\vec z - \vec w|^2}\)^{\D} \sim \pi^{\frac{d}{2}}\frac{\Gamma\(\D - \frac{d}{2}\)}{\Gamma({\D})} y^{d-2\D} \delta^d(\vec z -\vec w) + \frac{1}{|\vec z - \vec w|^{2\D}}, \  y\rightarrow 0
\ee
In our case we have to expand the expression 
\be
\frac{1}{\(q(\vec w)\cdot x\)^{\D}} = \frac{1}{\( u q^0 + 2R | \vec w - \vec z |^2 \)^{\D}} = \frac{1}{(2R)^{\D}} \frac{1}{\(y^2 + |\vec w - \vec z|^2\)^{\D}}
\ee
where now
\be
y^2 = \frac{uq^0}{2R},  \  q^0 = 1+ \vec w^2,  \  R\rightarrow \infty
\ee
After simple algebra we get the following large $R$ expansion
\be\label{contact}
\begin{gathered}
\frac{1}{\(q(\vec w)\cdot x\)^{\D}} \sim \pi^{\frac{d}{2}}\frac{\Gamma\(\D - \frac{d}{2}\)}{\Gamma({\D})} \frac{(uq^0)^{\frac{d}{2}-\D}}{(2R)^{\frac{d}{2}}} \delta^d(\vec z -\vec w) \\
+ \frac{1}{(2R)^{\D}} \frac{1}{|\vec z - \vec w|^{2\D}}
\end{gathered}, 
\  R\rightarrow\infty
\ee
Note that since we are considering operators on the unitary principal series for which $\D = \frac{d}{2} + i\lambda$ both the contact and the non-contact terms in \eqref{contact} are of the same order as $R\rightarrow\infty$. 

Now substituting the expansion in \eqref{massless} the final answer can be written schematically as
\be\label{contactshadow}
\boxed{
G_{N}(\{x_i\rightarrow\infty,\delta_i\}) \sim \sum_{n+\tilde n = N, \{\Delta_i, \D_{\tilde i}\}} R^{- \frac{nd}{2}-\sum_{i=\tilde1}^{\tilde n} \D_{\tilde i}} < \phi_{d-\D_1}^{\delta_1}(\vec z_1)\cdots \phi_{d-\D_{ n}}^{\delta_{n}}(\vec z_n)\tilde\phi_{\D_{\tilde 1}}^{\delta_{\tilde 1}}(\vec z_{\tilde1})\cdots  \tilde\phi_{\D_{\tilde n}}^{\delta_{\tilde n}}(\vec z_{\tilde n})>}
\ee
where $n$ is the number of conformal primaries and $\tilde n$ is the number of shadow conformal primaries. 

So we can see that the extrapolate limit of the time ordered Green's functions can be expressed as a sum of Celestial correlators of both shadow and non-shadow operators. The right hand side of \eqref{contactshadow} contains contact terms. To see this let us consider the four point function in four dimensions. The term with no shadow insertion is the conventional celestial amplitude which is known to be a distribution containing $\delta(z-\bar z)$ where $z$ is the conformal cross-ratio of the four points. On the other hand the term with only shadow insertions is expected to be analytic and captures the boundary four point function of \cite{Jain:2023fxc}. If we follow the standard procedure then we should discard the contact term from the four point function and keep only the celestial shadow correlator on the RHS of \eqref{contactshadow}. This will be the right thing to do for a standard quantum field theory but it is not clear if Celestial CFT is a standard CFT or not. Even in a standard CFT contact terms often carry universal information. For example, the two point function of the trace of the stress tensor in a $2-D$ CFT is a pure contact term, $<T_{z\bar z}(z) T_{z\bar z}(0)>\sim c \partial\bar\partial \delta^2(z)$, where $c$ is the central charge. This is a universal contact term and we cannot set it to zero without setting $c=0$. There are many such examples. 

The current situation deserves better understanding\footnote{See \cite{Jorstad:2024yzm,Kulp:2024scx} for recent works in this direction.} and we hope to return to this problem in future. 

\section{Acknowledgement}
I would like to thank the referee for very useful comments.
The work of SB is partially supported by the  Swarnajayanti Fellowship (File No- SB/SJF/2021-22/14) of the Department of Science and Technology and SERB, India.

\end{document}